\documentclass[3p]{elsarticle}
\usepackage{epstopdf}
\usepackage{amsmath, amsfonts, amssymb}
\usepackage{amsthm}
\usepackage{yhmath}
\usepackage{epsf}
\usepackage{amsfonts}
\usepackage{stmaryrd}
\usepackage{amssymb}
\usepackage{leftidx}
\usepackage{mathtools}
\usepackage{placeins}
\usepackage{booktabs}
\usepackage{enumitem}
\usepackage{caption}
\usepackage{tabu}
\usepackage{multirow}
\usepackage{stackengine}
\usepackage[utf8]{inputenc}
\usepackage[english]{babel}
\usepackage{bm}
\usepackage{array}
\usepackage{multirow}
\usepackage{graphicx} % Required for inserting images
\usepackage[T1]{fontenc}
\usepackage[utf8]{inputenc}
\usepackage{amssymb}
\usepackage{subfiles} 
\usepackage{subcaption}
\usepackage{multicol}
\usepackage{bm}
\usepackage{graphicx}
\usepackage[export]{adjustbox}
\usepackage[font=footnotesize]{caption}
\usepackage{float} %picture placement
\usepackage{placeins}%floatbarrier
\usepackage{longtable}

\usepackage{amsmath,stmaryrd,footmisc}
\usepackage{esint}
\usepackage{placeins}
\usepackage{algorithm}
\usepackage{algpseudocode}
\usepackage{esvect}
\MakeRobust{\vv}

\usepackage{epstopdf}
\usepackage{natbib}
\usepackage{nohyperref}
\usepackage{graphicx}
 \usepackage{setspace}
\usepackage{color}
\usepackage{verbatim}
 \usepackage[usenames,dvipsnames]{pstricks}
 \usepackage{epsfig}
 \usepackage{pst-grad} % For gradients
 \usepackage{pst-plot} % For axes
\usepackage{etoolbox}
\usepackage{mathtools}
\preto\subequations{\ifhmode\unskip\fi}

\usepackage{enumerate}
\usepackage{enumitem}
\usepackage{amsmath}
\usepackage{placeins}
\usepackage{siunitx}
\usepackage{cleveref}
\usepackage{empheq, esdiff}

\usepackage{physics}
\def\nten#1{\mathbf{#1}}

\def\sss{\sigma_{\texttt{ss}}}
\def\sbs{\sigma_{\texttt{bs}}}
\def\sigbar{\bar{\bm{\sigma}}}
\def\sigbarsi{\bar{\bm{\sigma}}_{\texttt{s}i}}

\def\Wbarts{\overline{W}_{\texttt{ts}}}
\def\Wbarss{\overline{W}_{\texttt{ss}}}
\def\Wbarcs{\overline{W}_{\texttt{cs}}}
\def\Wbarbs{\overline{W}_{\texttt{bs}}}

\def\bfu{{\bf u}}

\def\bfN{{\bf N}}

\def\bfX{{\bf X}}
\def\bfF{{\bf F}}

\def\eps{\varepsilon}

\def\sig{\sigma}

\def\e0{\varepsilon_0}

\def\s0{\sigma_0}

\def\sts{\sigma_{\texttt{ts}}}
\def\scs{\sigma_{\texttt{cs}}}

\def\de{\delta^\varepsilon}
\def\ce{c_{\texttt{e}}}

\DeclareMathAlphabet{\mathsfit}{T1}{\sfdefault}{\mddefault}{\sldefault}
\SetMathAlphabet{\mathsfit}{bold}{T1}{\sfdefault}{\bfdefault}{\sldefault}

\theoremstyle{definition}
\newtheorem{theorem}{Theorem}

\newtheorem{remark}[theorem]{Remark}
\newtheorem*{definition}{Definition}

\long\def\symbolfootnote[#1]#2{\begingroup%
\def\thefootnote{\fnsymbol{footnote}}\footnote[#1]{#2}\endgroup}

\begin{document}
\begin{frontmatter}

\title{On the construction of explicit analytical driving forces for crack nucleation in the phase field approach to brittle fracture with application to Mohr-Coulomb and Drucker-Prager strength surfaces}

\author[purdue,MIT]{S. Chockalingam\corref{cor1}}
\ead{chocka94@gmail.com}

\address[purdue]{Department of Mechanical Engineering, Purdue University, West Lafayette, IN 47907, USA }
\address[MIT]{Department of Aeronautics and Astronautics, Massachusetts Institute of Technology, Cambridge, MA 02139, USA}

\cortext[cor1]{Corresponding author}

\begin{abstract}
A series of recent papers have modified the classical variational phase-field fracture models to successfully predict both the nucleation and propagation of cracks in brittle fracture under general loading conditions. This is done through the introduction of a consistent crack nucleation driving force in the phase field governing equations, which results in the model being able to capture both the strength surface and fracture toughness of the material. This driving force has been presented in the literature for the case of Drucker-Prager strength surface and specific choice of stress states on the strength surface that are captured exactly for finite values of the phase field regularization length $\eps$. Here we present an explicit analytical expression for this driving force given a general material strength surface when the functional form of the strength locus is linear in the material parameter coefficients. In the limit $\eps \to 0$, the formulation reproduces the exact material strength surface and for finite $\eps$ the strength surface is captured at any $n$ `distinct' points on the strength surface where $n$ is the minimum number of material coefficients required to describe it. The presentation of the driving force in the current work facilitates the easy demonstration of its consistent nature. Further, in the equation governing crack nucleation, the toughness in the classical models is shown to be replaced by an effective toughness in the modified theory, that is dependent on the stress. The derived analytical expressions are verified via application to the widely employed Mohr-Coulomb and Drucker-Prager strength surfaces.

\keyword{Phase-field regularization; Crack nucleation driving force; Fracture nucleation; Mohr-Coulomb; Drucker-Prager;}
\endkeyword

\end{abstract}

\end{frontmatter}

\section{Introduction}

A phase field approximation of the variational sharp theory of brittle fracture \citep{Francfort98} was put forth in \cite{Bourdin00,Bourdin08}, the theory therein and all related models\footnote{Specifically phase field models of fracture that $\Gamma$ - converge to the variational theory of brittle fracture by \cite{Francfort98}, see \cite{lopez2025classical}.} will be referred to as classical (variational) phase field models in this manuscript. A modification to the classical phase field models was proposed in \cite{KFLP18,KBFLP20} wherein a crack nucleation driving force is introduced to the governing equations that marries the strength surface and toughness of the material such that the new framework is able to successfully predict both the nucleation and propagation of cracks, the former of which is not possible with the classical models \citep{lopez2025classical,kamarei2025nucleation}. A host of papers have since successfully demonstrated the ability of this general phase field theory of brittle fracture to predict fracture and failure in a wide range of materials and loading conditions \citep{KRLP18,KLP20,KLP21,KRLP22,fei2023phase,jia2023controlling,KLDLP24,KKLP24,liu2024emergence,liu2024effects}. It has also been cast in a two-step variational form \citep{larsen2024} and as a phase transition theory within the framework of configurational forces \citep{KRLP18}.  While a general roadmap for the construction of the crack nucleation driving force was outlined in \cite{KBFLP20}, a systematic establishment of the derivation of an explicit and consistent expression for a general strength surface is lacking. We tackle this in this manuscript by developing an explicit expression for a consistent crack nucleation driving force given a general material strength surface that is linear in its material coefficient parameters. 
The current work greatly simplifies the presentation of the driving force and facilitates the easy demonstration of its consistency (such that the material strength surface is exactly predicted in the limit of vanishing phase field regularization length) as well as a potential physical interpretation of the theory. We validate the developed expressions by applying them to the modeling of Mohr-Coulomb and Drucker-Prager strength surfaces, the former of which has not been previously modelled using a crack nucleation driving force but is a fundamentally important strength surface in the modeling of both brittle and ductile fracture \citep{yu2002advances,BaiWierzbicki2010}. Thus the expressions herein should facilitate any potential extension of the theory to predict crack nucleation in ductile fracture. While expressions for the Drucker-Prager strength surface have previously been presented \citep{KBFLP20,KRLP22,KKLP24}, here we show results for a different constitutive choice of strength calibration as will become clear later.\\

The manuscript is organized as follows, In \Cref{sec:theory}, we begin with a quick overview of the modified phase field theory and the notion of a consistent crack nucleation driving force. We then proceed to develop an explicit expression for this driving force that satisfies these requirements. Finally the developed expressions are validated by using them to model the Drucker-Prager and Mohr-Coulomb strength surfaces in \Cref{sec:results}. Brief concluding remarks are provided in \Cref{sec:conclusions}.

\section{Theory}
\label{sec:theory}

\subsection{Overview of the modified phase field theory}
\label{subsec:overview}

We begin with a quick overview of the modified phase field theory. Consider a structure made of an isotropic elastic brittle material occupying an open bounded domain $\mathrm{\Omega}\subset \mathbb{R}^3$ with boundary $\partial\mathrm{\Omega}$ and outward unit normal $\nten{N}$, in its undeformed and stress-free configuration at time $t=0$. At a later time $t \in (0, T]$, due to an externally applied displacement $\tilde{\bfu}(\bfX, t)$ on a part $\partial\mathrm{\Omega}_\mathcal{D}$ of the boundary and a traction $\tilde{\textbf{t}}(\bfX,t)$ on the complementary part $\partial\mathrm{\Omega}_\mathcal{N}=\partial\mathrm{\Omega}\setminus \partial\mathrm{\Omega}_\mathcal{D}$, the material points in the structure described by position vector $\nten{X}$ are displaced to a new position $\nten{x} = \nten{X} + \nten{u}(\nten{X},t)$. In addition to the displacement, the applied boundary conditions might also nucleate and propagate cracks in the structure. The cracks are described in a regularized way by the phase field variable $v=v(\bfX,t)$ taking values in $[0,1]$. Regions of pristine material are described by $v=1$ whereas $v<1$ identifies regions of the material that have been fractured. The deformation gradient is defined as $\nten{F} = \nten{I} + \nabla \nten{u}$ and the strain energy function of the material is described by $\nten{W}(\nten{F})$. In addition to the strain energy, the material is described by a fracture toughness or critical energy release rate $G_c$ and a material strength surface described below.\\

\noindent \textbf{Material strength surface :} When a macroscopic piece of an elastic brittle material is subjected to a state of monotonically increasing and spatially uniform (but arbitrary) stress, fracture will nucleate at a critical value of the applied stress. The set of all such critical stresses defines a surface in stress space. This surface is referred to as the strength surface of the material and is considered a material property (save for stochasticity) in the macroscopic theory here \citep{KBFLP20}. It is a macroscopic manifestation of the presence of microscopic defects. The material strength surface is described as follows
\begin{equation}
    \mathcal{F} \equiv \hat{F}(\bm{\sigma}) = 0 \label{eq:firstmaterial_ss}
\end{equation}
where $\bm{\sigma}$ is any suitably defined stress measure. For example, the Biot stress has emerged as a choice for elastomeric materials \citep{KLP21,KKLP24}, defined as $\bm{\sigma} = \left(\nten{S}^T\nten{R} + \nten{R}^T\nten{S}\right)/2$ where $\nten{S}$ is the first Piola–Kirchhoff stress and the deformation gradient $\nten{F}$ admits a right polar decomposition $\nten{F} = \nten{R} \nten{U}$ in terms of the rigid rotation tensor $\nten{R}$ and the right stretch tensor $\nten{U}$. Further, for any stress state $\bm{\sigma}$ before attainment of strength failure (not on the locus $\mathcal{F}$), we assume $\hat{F}({\bm{\sigma}}) < 0$ and that $\hat{F}({\bm{\sigma}}) > 0$ is in violation of the strength of the material.

\begin{remark}     
 Let ${\bm{\sigma}} =  |{\bm{\sigma}}|~ \nten{n}$ where $|{\bm{\sigma}}| =\sqrt{{\bm{\sigma}} : {\bm{\sigma}}}$. For proportional loadings, that is for a fixed $\nten{n}$, $\hat{F}$ is expected to be an increasing function of the magnitude of the stress ($|{\bm{\sigma}}|$). Thus $\hat{F}$ attains its minimum value at ${\bm{\sigma}} = \nten{0}$.
\end{remark}

\begin{remark}  
The nucleation of cracks in a pristine material for uniform applied stress in the absence of large cracks is described by the material strength whereas the propagation of large cracks is described by the fracture toughness $G_c$ through the Griffith theory of fracture. In all other scenarios, fracture is mediated by an interplay of these two limits and the modified phase field theory is a unified macroscopic continuum description of the fracture process.
\end{remark}
 
 In its latest form, the governing equations of the modified phase field theory introduced by \cite{KFLP18,KBFLP20}
 can be written as follows \citep{larsen2024,KKLP24,lopez2025classical}; the displacement field $\bfu_k(\bfX)=\bfu(\bfX,t_k)$, and phase field $v_k(\bfX)=v(\bfX,t_k)$ at any material point $\bfX\in\overline{\mathrm{\Omega}} = \mathrm{\Omega} \cup \partial\mathrm{\Omega}$ and discrete time $t_k\in\{0=t_0,t_1,...,t_m,$ $t_{m+1},...,$ $t_M=T\}$ are determined by the following system of coupled partial differential equations:
\begin{subequations}
\begin{align}
 &{\rm Div}\left[v_{k}^2\dfrac{\partial W}{\partial \bfF}(\bfF(\bfu_{k}))\right]={\bf0},~ \bfX\in\mathrm{\Omega},\\
&\bfu_{k}=\tilde{\bfu}(\bfX,t_{k}),~ \bfX\in\partial  \mathrm{\Omega}_{\mathcal{D}},\\
& \left[v_{k}^2\dfrac{\partial W}{\partial \bfF}(\bfF(\bfu_{k}))\right]\bfN=\tilde{\textbf{t}}(\bfX,t_{k}),\quad \bfX\in\partial \mathrm{\Omega}_{\mathcal{N}}
 \end{align} \label{BVP-u-theory}
\end{subequations}
and\\
\small
\begin{subequations}
\begin{align}
&\dfrac{3}{4} \varepsilon \, \de \,  G_c \triangle v_{k}=2 v_{k} W(\bfF(\bfu_{k}))-c_\texttt{e}(\bfX,t_{k})- \dfrac{3}{8}  \dfrac{\de \, G_c}{\varepsilon},\quad
 \mbox{if } v_{k}(\bfX)< v_{k-1}(\bfX),\quad \bfX\in\mathrm{\Omega} \label{BVP-v-theory_main}\\
&\dfrac{3}{4} \varepsilon \, \de \,  G_c \triangle v_{k}\geq2 v_{k} W(\bfF(\bfu_{k}))-c_\texttt{e}(\bfX,t_{k})- \dfrac{3}{8} \dfrac{\de \, G_c}{\varepsilon},\quad  
 \mbox{if } v_{k}(\bfX)=1\; \mbox{ or }\; v_{k}(\bfX)= v_{k-1}(\bfX)>0,\quad \bfX\in\mathrm{\Omega} \label{BVP-v-theory_secondary}\\
&\normalsize v_{k}(\bfX)=0,\quad \mbox{ if } v_{k-1}(\bfX)=0,\quad \bfX\in\mathrm{\Omega}
\\
&\nabla v_{k}\cdot\bfN=0,\quad \bfX\in \partial\mathrm{\Omega} 
   \end{align} \label{BVP-v-theory}
\end{subequations}
\normalsize with example initial conditions $\bfu(\bfX,0)\equiv\textbf{0}$ and $v(\bfX,0)\equiv1$, where $\triangle(\cdot) \equiv \text{Div}(\nabla(\cdot))$ and $\varepsilon>0$ is a regularization or localization length.  The term $c_\texttt{e}(\bfX,t)$ is the crack nucleation driving force that contains information about material strength and $\de$ is a nonnegative coefficient whose specific constitutive prescription depends in turn on the particular form of $\ce$. Note that setting $\de = 1$ and $\ce=0$ reduces the formulation to the classical phase field theories which can successfully model large crack propagation physics since they converge to the Griffith theory of fracture in the limit of $\eps \to 0$. Thus the introduction of $\ce$ in the theory is meant to capture the material's strength surface while a suitably tuned coefficient $\de$, that is a function of $\eps$, preserves the ability of the phase field theory to model large crack fracture physics. The Piola stress is given by the equation ${\nten{S}} = v^2 \pdv{W(F)}{\nten{F}}$ from which the Biot stress (or other suitable stress measure) $\bm{\sigma}$ can be obtained as well. Thus we note that the strain energy function can be written as $W(\bm{\sigma}(\nten{F},v),v).$\\

\noindent \textbf{The phase-field strength surface :} For uniform and monotonically increasing applied stresses in the presence of no cracks in the material ($v=1$ everywhere), under the assumption of suitable localization of $v$, the above phase field theory predicts crack nucleation due to strength failure according to the following strength surface ${\mathcal{F}}^\eps$ \citep{KBFLP20},
\begin{equation}
{\mathcal{F}}^\eps \equiv  2 W(\bm{\sigma})-c_\texttt{e} - \omega_\eps = 0 , \qquad \omega_\eps = \dfrac{3}{8}  \dfrac{\de \, G_c}{\varepsilon} \label{eq:First_phase_ss_general}
\end{equation}
where $W(\bm{\sigma}) = W(\bm{\sigma}(\nten{F},v=1),v=1)$. Note that setting $v_k=1$ in \cref{BVP-v-theory_main} yields \cref{eq:First_phase_ss_general}, however see \cite{pham2011gradient} and Appendix of \cite{KBFLP20} for a discussion on stability and localization. We shall henceforth refer to \cref{eq:First_phase_ss_general} as the phase field strength surface. Now we define the notion of a consistent crack nucleation driving force.

\begin{definition}
A consistent crack nucleation driving force $\ce$ is one such that the phase field strength surface, \cref{eq:First_phase_ss_general}, reduces to the material strength surface, \cref{eq:firstmaterial_ss}, in the limit of vanishing phase field regularization length ($\eps \to 0$).
\end{definition}

It has been demonstrated through several numerical studies \citep{KBFLP20,KLP20,KRLP22,KLDLP24,KKLP24,liu2024effects} that the introduction of a consistent $\ce$ that is a function of the true stress\footnote{Note for instance that $\ce$ cannot be prescribed as function of $\bm{\sigma}(\nten{F},v = 1 )$ if large crack propagation physics is to be preserved.}, $\bm{\sigma} = \bm{\sigma}(\nten{F},v)$,  will preserve Griffith large crack physics in an arbitrary boundary value problem with large pre-existing crack as long the following is done - the coefficient $\delta^\eps$ is numerically selected (for a given $\eps$) such that the theory accurately predicts nucleation from a large pre-existing crack for \textit{any one} boundary value problem governed by the Griffith theory of fracture. Due to its simplicity of analysis, the ``pure-shear'' fracture test introduced by \cite{rivlin1953rupture} is often a suitable choice for elastomers, or the single edge notch test or middle crack tension test in the case of linear elastic fracture mechanics. We note that \cite{KKLP24} provided an explicit phenomenological expression for $\delta^{\eps}$ following extensive numerical simulations that suggests the following functional form
\begin{equation}
    \delta^\eps = \frac{2}{5} + \frac{G_c}{f~\eps} \label{eq:suggested_delta_epsilon}
\end{equation}
where $f$ is a scalar function of the strength and elastic material parameters and has the dimensions of stress. A definitive comprehensive understanding of how the theory is able to preserve large crack physics through  suitably tuned $\de$ despite perturbation of the governing equations of classical phase field theory by the nucleation driving force is still lacking, however we do not attempt to address this in this manuscript. Instead, we outline explicit analytical expressions for a consistent $\ce$ given a general material strength surface $\mathcal{F}$ satisfying certain requirements whose form we discuss in the following section.

\subsection{Form of the considered material strength surface}

Consider the following general form of the material strength surface written in terms of the dimensionless stress ($\bar{\bm{\sigma}})$ and $n$ dimensionless material constants $\vv{\beta}$,
\begin{equation}
    \mathcal{F} \equiv F(\bar{\bm{\sigma}}, \vv{\beta}) =  g(\bar{\bm{\sigma}} , \vv{\beta}) -1 = 0, \quad \bar{\bm{\sigma}} = \dfrac{\bm{\sigma}}{\sigma^*}, \quad \vv{\beta} = [\beta_i] \quad  i =1,2,3,...,n \label{eq:generalform_target_ss}
\end{equation}
where $\sigma^*$ is a suitably chosen characteristic stress. We will assume here that the function $g$ is linear in the parameters $\vv{\beta}$. For example, below we write two widely employed strength surfaces - Mohr-Coulomb and Drucker-Prager, in the above form using two constants ($n = 2$) for both.

\begin{remark}
     Typically $g(\nten{0},\vv{\beta}) = 0$ so that $F(\nten{0}, \vv{\beta}) = -1$, which are also their respective minimum values.
\end{remark}\vspace{0.3 cm}

\noindent \textbf{Mohr-Coulomb strength surface :} We can write the Mohr-Coulomb (M-C) strength surface in the form in \cref{eq:generalform_target_ss}  as follows
\begin{equation}
   \mathcal{F}_{\texttt{MC} }\equiv {F}_{\texttt{MC}}(\bar{\bm{\sigma}}, \beta_1, \beta_2) =  g_{\texttt{MC}}(\bar{\bm{\sigma}} , \beta_1, \beta_2) -1 = 0, \quad g_{\texttt{MC}}(\bar{\bm{\sigma}}, \beta_1, \beta_2) = \beta_1 \bar{{\sigma}}_{\texttt{max}} + \beta_2 \bar{{\sigma}}_{\texttt{min}}, \label{eq:generalform_MC} 
\end{equation}
where $\bar{{\sigma}}_{\texttt{max}},\bar{{\sigma}}_{\texttt{min}}$ are the maximum and minimum principal values of $\sigbar$ respectively. Let $\sts$, $\scs$, and $\sss$ denote the uniaxial tensile, uniaxial compressive, and shear strengths of the material respectively. We make the choice $\sig^* = \sts$ in this manuscript. We can then write the dimensionless constants $\beta_1, \beta_2$ as follows
\begin{equation}
    \beta_1 = 1, \quad \beta_2 = -\alpha_{\texttt{tc}} \quad \text{where} \quad \alpha_{\texttt{tc}} = \frac{\sts}{\scs} = \frac{1}{\bar{\sigma}_{\texttt{cs}}}\label{eq:MC_betas}
\end{equation}
or as
\begin{equation}
    \beta_1 = 1, \quad \beta_2 = 1 - \alpha_{\texttt{ts}} \quad \text{where} \quad \alpha_{\texttt{ts}} = \frac{\sts}{\sss} = \frac{1}{\bar{\sigma}_{\texttt{ss}}}\label{eq:MC_betas2}
\end{equation}
More generally, the $n$ dimensionless parameters can be expressed as a function of any $n$ `distinct' dimensionless stress states on the strength surface as shown later in \cref{eq:beta_strengthrelation} (note that for our choice of $\sig^*$ here, the dimensionless tensile strength is unity).

\begin{remark}
    The Tresca surface is a special limit of the M-C surface when $\alpha_{\texttt{tc}} = 1$.
\end{remark}  \vspace{0.4cm}

\noindent \textbf{Drucker-Prager strength surface :}
We can write the Drucker-Prager (D-P) strength surface in the form in \cref{eq:generalform_target_ss}  as follows
\begin{equation}
   \mathcal{F}_{\texttt{DP}} \equiv {F}_{\texttt{DP}}(\bar{\bm{\sigma}}, \beta_1, \beta_2) =  g_{\texttt{DP}}(\bar{\bm{\sigma}}, \beta_1, \beta_2)-1, \quad g_{\texttt{DP}}(\bar{\bm{\sigma}}, \beta_1, \beta_2) = \beta_1 \bar{I}_1 + \beta_2 \sqrt{\bar{J}_2}, \label{eq:generalform_DP}
\end{equation}
where 
\begin{equation}
    \bar{I}_1 = \text{tr}(\bar{\bm{\sigma}}), \quad \bar{J}_2 = \frac{1}{2}\text{tr}(\bar{\bm{\sigma}}_D^2), \quad \bar{\bm{\sigma}}_D = \bar{\bm{\sigma}} - \frac{1}{3}\text{tr}(\bar{\bm{\sigma}}) \nten{I}
\end{equation} 
 The dimensionless constants $\beta_1, \beta_2$ can be written as
\begin{equation}
    \beta_1 = 1-\frac{\alpha_{\texttt{ts}}}{\sqrt{3}}, \quad \beta_2 = \alpha_{\texttt{ts}} \label{eq:DP_betas1} 
\end{equation}
or as 
\begin{equation}
    \beta_1 = {\alpha_{\texttt{tb}}}-1, \quad \beta_2 = \sqrt{3}\left(2-{\alpha_{\texttt{tb}}}\right), \quad \alpha_{\texttt{tb}} = \frac{\sts}{\sbs} = \frac{1}{\bar{\sigma}_{\texttt{bs}}} \label{eq:DP_betas2}
\end{equation}
where $\sbs$ is the tensile biaxial strength of the material and we have the relation $2{\alpha_{\texttt{tb}}} = 3- {\alpha_{\texttt{tc}}}$.

\begin{remark}
    The von-Mises surface is a special limit of the D-P surface when $\alpha_{\texttt{tc}} = 1$.
\end{remark}  

\subsection{Construction of the driving force $\ce$}

We now construct a consistent driving force $\ce$ for the material strength surface in \cref{eq:generalform_target_ss}. First, we write the phase field strength surface \cref{eq:First_phase_ss_general} in the following form using dimensionless terms,
\begin{equation}
 {\mathcal{F}}^\eps \equiv 2 \overline{W}^\eps - \bar{c}_{\texttt{e}}^\eps   - 1 = 0, \quad  \overline{W}^\eps = \dfrac{W}{\omega_\eps}, \quad \bar{c}_{\texttt{e}}^\eps = \dfrac{\ce}{\omega_\eps} \label{eq:general_dimlss_ss_pf}
\end{equation}
where $\eps$ as a superscript over an overline bar represents non-dimensionlization by the quantity $\omega_\eps$ (defined in \cref{eq:First_phase_ss_general}) which has the dimensions of stress (to distinguish from non-dimensionalization by $\sig^*$)
We can further define
\begin{equation}
    \bar{\omega}_\eps = \dfrac{\omega_\eps}{\sigma^*}, \quad  \overline{W} =  \frac{W}{\sigma^*}, \quad \overline{W}^\eps = \dfrac{\overline{W}}{\bar{\omega}_\eps} = \dfrac{8}{3} \dfrac{\eps}{\de} \dfrac{ 
 \sigma^*\overline{W}}{G_c} \label{eq:dimlssdefs}
\end{equation}
We note that $\overline{W}$ can be written as a function of the dimensionless stress so that $\overline{W} = \overline{W}(\bar{\bm{\sigma}})$ and thus $\overline{W}^\eps = \overline{W}^\eps(\bar{\bm{\sigma}})$.\\

We choose the following form for the dimensionless driving force $\bar{c}_{\texttt{e}}$ (recall the function $g$ from \cref{eq:generalform_target_ss}),
\begin{equation}
   { \bar{c}_{\texttt{e}}^\eps = -g(\bar{\bm{\sigma}}, \vv{{\beta}^\eps}),\quad \vv{\beta^\eps} = \vv{\beta} + \vv{\Delta {\beta}^\eps}} \label{eq:general_chosen_cebar}
\end{equation}
where $\beta_i^\eps,\Delta {\beta}_i^\eps$ (we indicate the components of a scalar array $\vv{y}$ as $y_i$) are $\eps$ dependent dimensionless scalars chosen such that 
\begin{equation}
    \Delta {\beta}_i^\eps \to 0 \quad \text{or equivalently} \quad \beta_i^\eps \to \beta_i, \quad \text{as} \quad \eps \to 0 , \qquad i =1,2,3,...,n\label{eq:delta_beta_limit}
\end{equation}
We will shortly prescribe $\Delta {\beta}_i^\eps$ that satisfy this requirement. Using \cref{eq:general_chosen_cebar} and \cref{eq:general_dimlss_ss_pf}$_3$ we arrive at the following expression for the driving force $\ce$
\begin{equation}
    \ce = -\omega_\eps g(\bar{\bm{\sigma}}, \vv{\beta} + \vv{\Delta {\beta}^\eps}) =  -\left(\dfrac{3}{8}  \dfrac{\de \, G_c}{\varepsilon}\right) g(\bar{\bm{\sigma}}, \vv{\beta} + \vv{\Delta {\beta}^\eps})  \label{eq:firstconstructed_ce}
\end{equation}\\

\noindent \textbf{Phase field strength surface for chosen $\ce$: }
Substituting the chosen driving force from \cref{eq:general_chosen_cebar} in \cref{eq:general_dimlss_ss_pf} yields the following form of the phase field strength surface
\begin{equation}
    {{\mathcal{F}}^\eps \equiv  2 \overline{W}^\eps(\bar{\bm{\sigma}}) + {F}(\bar{\bm{\sigma}}, \vv{\beta^\eps}) = 0 }\label{eq:final_general_ss_pf}
\end{equation}
Note from \cref{eq:dimlssdefs} that as $\eps \to 0$ we get $\overline{W}^\eps \to 0$ as long as $\eps/\de \to 0$ (or $\bar{\omega}_\eps \to \infty$). The suggested functional form of $\de$ in \cref{eq:suggested_delta_epsilon} satisfies this property and we will assume that the numerically calibrated $\delta^\eps$ satisfies the requirement that $\bar{\omega}_\eps \to \infty$ as $\eps \to 0$. Any mention of the limit $\bar{\omega}_\eps \to \infty$ in the manuscript can thus be considered the limit of $\eps \to 0$. Thus, using \cref{eq:delta_beta_limit} in \cref{eq:final_general_ss_pf}, we get
\begin{equation}
    {\mathcal{F}}^\eps \to \mathcal{F} \equiv F(\bar{\bm{\sigma}}, \vv{\beta}) = 0 \quad \text{as} \quad \eps \to 0
\end{equation}
i.e., the phase field strength surface reduces exactly to the material strength surface in the limit $\eps \to 0$ as long as \cref{eq:delta_beta_limit} is satisfied. Thus the prescription of any $\vv{\Delta {\beta}^\eps}$ that satisfies \cref{eq:delta_beta_limit} completes the construction of a consistent crack nucleation driving force, which we do next. \\

\noindent \textbf{Choice of $\vv{\Delta \beta^\eps}$ :}
While any $\vv{\Delta {\beta}^\eps}$ that satisfies \cref{eq:delta_beta_limit}, including the choice $\vv{\Delta {\beta}^\eps} = \vv{0}$, will satisfy the requirement of consistency of $\ce$ in the limit $\eps \to 0$, we note that in practice $\eps$ is a finite length parameter in simulations\footnote{It only needs to be selected to be smaller than the smallest material length scale built into \cref{BVP-u-theory,BVP-v-theory}, which arises from different units of the strain energy function, the strength, and toughness, see Appendix C in \cite{KLDLP24} and Appendix B in \cite{KKLP24}.}. Thus it does not suffice to merely satisfy the consistency requirement and we enforce the following additional requirement.\\ 

\noindent{\textbf{Finite $\bm{\eps}$ match requirement(*) for $\vv{\Delta \beta^\eps}$ :}}
   We choose $\vv{\Delta {\beta}^\eps}$ such that the phase field strength surface condition in \cref{eq:final_general_ss_pf} is satisfied at $n$ distinct (in a certain sense specified later) chosen stress states ($\bar{\bm{\sigma}}_{\texttt{s}i}$) that also satisfy the material strength surface condition in \cref{eq:generalform_target_ss}. Thus at these stress states, the phase field strength surface coincides with the material strength surface \textit{irrespective of the value of $\eps$}.\\

\noindent Ideally, it would be best practice to choose $\bar{\bm{\sigma}}_{\texttt{s}i}$ such that they cover the most dominant expected modes of strength failure in the problem at hand. The requirement (*) can be mathematically written as
\begin{equation}
 2 \overline{W}^\eps(\bar{\bm{\sigma}}_{\texttt{s}i}) + {F}(\bar{\bm{\sigma}}_{\texttt{s}i}, \vv{\beta^\eps}) = 0  \quad  \text{for} \quad i =1,2,3,... , n  \label{eq:surfacmatching}
\end{equation}
wherein $\bar{\bm{\sigma}}_{\texttt{s}i}$ satisfy
\begin{equation}
    {F}(\bar{\bm{\sigma}}_{\texttt{s}i}, \vv{\beta}) = 0 \quad  \text{for} \quad i =1,2,3,... , n  \label{eq:spec_strpoints}
\end{equation}
Using the linearity of the function $g$ in $\vv{\beta}$ and \cref{eq:spec_strpoints}, \cref{eq:general_chosen_cebar}$_2$, we can write
\begin{equation}
    {F}(\bar{\bm{\sigma}}_{\texttt{s}i}, \vv{\beta^\eps}) = {F}(\bar{\bm{\sigma}}_{\texttt{s}i}, \vv{\beta})  + \left[F_{i,\vv{\beta}}\right]^T  [\vv{\Delta {\beta}^\eps}] = \left[F_{i,\vv{\beta}}\right]^T  [\vv{\Delta \beta^\eps}], \qquad F_{i,\vv{\beta}} = \pdv{F_i}{\vv{\beta}} = \pdv{F(\bar{\bm{\sigma}}_{\texttt{s}i}, \vv{\tilde{\beta}})}{\vv{\tilde{\beta}}} \biggr\vert_{\vv{\tilde{\beta}} = {\vv{\beta}}} 
    \label{eq:matrixeq}
\end{equation}
where $F_{i,\vv{\beta}}$ are dimensionless constants\footnote{Note that when the function $g$ linear in $\beta$, $\pdv{F(\bar{\bm{\sigma}}_{\texttt{s}i}, \vv{\tilde{\beta}})}{\vv{\tilde{\beta}}} \biggr\vert_{\vv{\tilde{\beta}} = {\vv{\beta}}} = \pdv{F(\bar{\bm{\sigma}}_{\texttt{s}i}, \vv{{\beta}})}{\vv{{\beta}}}$ are dimensionless constants.}.
Using \cref{eq:matrixeq} in \cref{eq:surfacmatching}, we arrive at the following matrix systems of equations
\begin{equation}
\begin{bmatrix}
F_{1,_{\beta_1}} & F_{1,_{\beta_2}} & \cdots & F_{n,_{\beta_n}} \\
F_{2,_{\beta_1}} & F_{2,_{\beta_2}} & \cdots & F_{2,_{\beta_n}} \\
\vdots & \vdots & \ddots & \vdots \\
F_{n,_{\beta_1}} & F_{n,_{\beta_1}} & \cdots & F_{n,_{\beta_n}}
\end{bmatrix}
\begin{bmatrix}
\Delta {\beta}_1^\eps \\
\Delta {\beta}_2^\eps \\
\vdots \\
\Delta {\beta}_{n}^\eps
\end{bmatrix} =
-2\begin{bmatrix}
\overline{W}^\eps_1 \\
\overline{W}^\eps_2 \\
\vdots \\
\overline{W}^\eps_n
\end{bmatrix}, \quad \overline{W}_i^\eps =  \overline{W}^\eps(\bar{\bm{\sigma}}_{\texttt{s}i})
\end{equation}
which we represent as follows
\begin{equation}
\left[{\pdv{{\vv{F}}}{\vv{\beta}}}\right] \vv{\Delta \beta^\eps}     = -2\vv{\overline{W}^\eps}
\end{equation}
Thus we can solve for $\vv{\Delta {\beta}^\eps}$ as
\begin{equation}
  {  \vv{\Delta \beta^\eps} = -\left[\pdv{\vv{F}}{\vv{\beta}}\right]^{-1}\frac{2\vv{\overline{W}}}{\bar{\omega}_\eps} = -\left(\dfrac{16}{3} \dfrac{\eps}{\de} \dfrac{ 
 \sigma^*}{G_c}\right)\left[\pdv{\vv{F}}{\vv{\beta}}\right]^{-1}\vv{\overline{W}} }, \quad \overline{W}_i =  \overline{W}(\bar{\bm{\sigma}}_{\texttt{s}i}) \label{eq:deltabetaeps_sol}
\end{equation}
Note that apart from the term in flower brackets there is no other dependence of $\vv{\Delta \beta^\eps}$ on $\eps$ and thus $\vv{\Delta \beta^\eps} \to \vv{0}$ as $\eps \to 0$ (or as $\bar{\omega}_\eps \to \infty$). Hence \cref{eq:delta_beta_limit} is satisfied and the constructed driving force in \cref{eq:firstconstructed_ce} is consistent.\\ 

\begin{remark}
    We can now clarify what it means for two strength states $\bar{\bm{\sigma}}_i$ and $\bar{\bm{\sigma}}_j$ to be distinct - it is the requirement that $F_{i,\vv{\beta}}$ and $F_{j,\vv{\beta}}$ are linearly independent. For $\left[\pdv{\vv{F}}{\vv{\beta}}\right]^{-1}$ to exist, the $n$ chosen strength states $\bar{\bm{\sigma}_i}$ need to be distinct so that the rows of the matrix $\left[\pdv{\vv{F}}{\vv{\beta}}\right]$ are linearly independent. 
\end{remark}  

\begin{remark}
    Note that for a material strength surface that is not linear in its material coefficients we can still satisfy the finite $\eps$ match requirement(*) for $\vv{\Delta \beta^\eps}$ by solving \cref{eq:surfacmatching}
for $\vv{\beta^\eps}$
and the solution in \cref{eq:deltabetaeps_sol} can be considered as a potential first order approximation that might capture the material strength surface in the limit $\eps \to 0$. However we are unable to mathematically establish the consistency of such a solution due to the reason that $F(\bar{\bm{\sigma}}_{\texttt{s}i}, \vv{\beta^\eps}) = {F}(\bar{\bm{\sigma}}_{\texttt{s}i}, \vv{\beta})$ at $n$ distinct strength locations does not necessitate  $F(\bar{\bm{\sigma}}, \vv{\beta^\eps}) = {F}(\bar{\bm{\sigma}}, \vv{\beta})$ for any $\bar{\bm{\sigma}}$ in the limit $\varepsilon \to 0$.
\end{remark} 

Further, using \cref{eq:spec_strpoints,eq:generalform_target_ss} and the linearity of $g$ in $\vv{\beta}$, we can write
\begin{equation}
  \vv{\beta}  = \left[\pdv{\vv{F}}{\vv{\beta}}\right]^{-1} \vv{1} \quad \text{where }\quad \vv{1} = [ 1\ 1  \hdots 1]^T \label{eq:beta_strengthrelation}
\end{equation}
This gives us the dimensionless parameters $\vv{\beta}$ in terms of $n$ distinct chosen dimensionless strength states $\bar{\bm{\sigma}}_{\texttt{s}i}$. Using \cref{eq:beta_strengthrelation,eq:deltabetaeps_sol}, we can write the following expression for $\vv{\beta^\eps}$
\begin{equation}
    \boxed{  \vv{ \beta^\eps} = \left[\pdv{\vv{F}}{\vv{\beta}}\right]^{-1}\left(\vv{1}-\frac{2\vv{\overline{W}}}{\bar{\omega}_\eps} \right) }
\end{equation}

\subsection{Final expressions and discussion}

Thus the final explicit analytical form of the driving force $\ce$ to capture the material strength surface $\mathcal{F}$ for all $\bar{\bm{\sigma}}$ in the limit $\eps \to 0$, and exactly at $n$ distinct states stress $\bar{\bm{\sigma}}_{\texttt{s}i}$ for any $\eps$, is given by
\begin{equation}
  \mathcal{F} \equiv F(\bar{\bm{\sigma}}, \vv{\beta}) = g(\bar{\bm{\sigma}}, \vv{\beta}) -1 = 0 \Rightarrow \boxed{   \ce =  -\left(\dfrac{3}{8}  \dfrac{\de \, G_c}{\varepsilon}\right) g(\bar{\bm{\sigma}}, \vv{\beta} + \vv{\Delta {\beta}^\eps})} \quad \text{where}
\end{equation}
\begin{equation}
  \boxed{  \vv{\Delta \beta^\eps}  = -\left(\dfrac{16}{3} \dfrac{\eps}{\de} \dfrac{ 
 \sigma^*}{G_c}\right)\left[\pdv{\vv{F}}{\vv{\beta}}\right]^{-1}\vv{\overline{W}} }, \quad \overline{W}_i =  \overline{W}(\bar{\bm{\sigma}}_{\texttt{s}i}), \quad \pdv{F_i}{\vv{\beta}} = \pdv{F(\bar{\bm{\sigma}}_{\texttt{s}i}, \vv{\beta})}{\vv{\beta}}, \quad {F}(\bar{\bm{\sigma}}_{\texttt{s}i}, \vv{\beta}) = 0
\end{equation}
The driving force yields a phase field strength surface  the form
\begin{equation}
   \boxed{{\mathcal{F}}^\eps \equiv  2 \overline{W}^\eps(\bar{\bm{\sigma}}) + {F}(\bar{\bm{\sigma}}, \vv{\beta^\eps}) = 0 } \quad \text{or} \quad \boxed{{\mathcal{F}}^\eps \equiv  2 {W}({\bm{\sigma}}) - \dfrac{3}{8}  \dfrac{ \, \hat{G}^\eps_c} {\varepsilon} = 0 } \quad \text{where} \quad \boxed{\hat{G}_c^\eps = -\de F(\bar{\bm{\sigma}}, \vv{\beta^\eps}) G_c}
\end{equation}
and the following form of the key governing equation \cref{BVP-v-theory_main} for the phase field variable when fracture is progressing, 
\begin{equation}
  \dfrac{3}{4} \varepsilon \, \de \,  G_c \triangle v_{k} =  2 v_k  W - \dfrac{3}{8}  \dfrac{ \, \hat{G}^\eps_c} {\varepsilon}
 \end{equation}
Note that in the classical phase field theories $\hat{G}_c^\eps$ would  be replaced by the fracture toughness $G_c$ instead. Thus it is seen that for crack nucleation, the toughness $G_c$ has been replaced by an effective toughness $\hat{G}_c^\eps$ that depends on the stress state. The  effective toughness is zero at initiation of strength failure\footnote{ The effective toughness exactly vanishes at the material strength surface for $\eps \to 0$ or at chosen $\sigbarsi$ for any $\eps$.} (since $F = 0$) and attains its maximum possible value of $\de G_c$ in the cracked state $v=0$ (since $\bar{\bm{\sigma}}(v=0) = \nten{0}$ and $F(\nten{0},\vv{\beta^\eps}) = -1$). This has interesting parallels to the cohesive zone models for fracture wherein the surface energy density or effective toughness usually starts out at zero and increases with the crack separation (which is related to the traction/stress at the crack surface), attaining a maximum value of $G_c$ \citep{Bourdin08}. The consequences of the notion of effective toughness and potential physical interpretation of the theory will be explored in future work. Nevertheless its emergence begs the question if the term $G_c$ on the left hand side of \cref{BVP-v-theory_main,BVP-v-theory_secondary} should also be potentially replaced by $\hat{G}_c^\eps$ for a more physically consistent notion of the modified phase field theory.\\ 

\noindent \textbf{Remark 8.}       Sometimes a correction factor can be introduced to the driving force \citep{KRLP22,KKLP24} to improve the phase field strength failure prediction (for finite $\eps$) in domains of stress space where fracture is not expected  (so that you do not have to worry about the correction disturbing the large crack propagation behaviour) such as when $\bar{I}_1<0$. Accordingly $\ce$ is prescribed as\footnote{Note that in the literature $h(\bar{I}_1)$ has been introduced as $\frac{1}{2}\left(1 - \frac{\sqrt{\bar{I}_1^2}}{\bar{I}_1}\right)$, but such a function is not defined for $\bar{I}_1 = 0$.}
\begin{equation}
  \ce =  -\omega_\eps g\left(\bar{\bm{\sigma}}, \vv{\beta^\eps}\right) + 2 v {W}(\nten{F}) h(\bar{I}_1), \quad h(\bar{I}_1) = \begin{cases}
      0 & \bar{I}_1 \ge 0 \\
      1 & \bar{I}_1 < 0
  \end{cases}
\end{equation}
so that the phase field strength surface becomes
\begin{equation}
    {\mathcal{F}}^\eps \equiv   2(1-h(\bar{I}_1)) \overline{W}^\eps(\bar{\bm{\sigma}}) + {F}(\bar{\bm{\sigma}}, \vv{\beta^\eps}) = 0 \label{eq:I1_correction}
\end{equation}
which reduces back to the usual form in \cref{eq:final_general_ss_pf} for loading states with $\bar{I}_1 >0$. Following the same solution process, we get the following solution for $\vv{\Delta \beta^\eps}$ instead
\begin{equation}
    \vv{\Delta \beta^\eps} = -\left[\pdv{\vv{F}}{\vv{\beta}}\right]^{-1}\frac{2\vv{\overline{W}'}}{\bar{\omega}_\eps}, \quad \overline{W}_i' = (1-h(\bar{I}_1(\sigbarsi))) \overline{W}_i 
\end{equation}
Thus if the chosen strength states $\sigbarsi$ are all such that $\bar{I}_1(\sigbarsi) \ge 0$, then the expressions for the coefficients $\vv{ \beta^\eps}$ will remain unchanged by the introduction of the correction term.
\\

This completes the development of the theory, we show application to the Mohr-Coulomb and Drucker-Prager material strength surfaces in the following section.

\section{Application of theory}
\label{sec:results}

We will now use our solution to construct $\ce$ for the M-C and D-P surfaces for different choices of $\bar{\bm{\sigma}}_{\texttt{s}i}$ and demonstrate that they satisfy both consistency and finite $\eps$ match requirement(*). For $n=2$, our solution for $\vv{\Delta \beta^\eps}$ is given by
\begin{equation}
   \begin{bmatrix}
        \Delta \beta^\eps_1\\
        \Delta \beta^\eps_2
    \end{bmatrix} = -\frac{2 }{\bar{\omega}_\eps \Lambda} \begin{bmatrix}
         F_{2,\beta_2}\overline{W}_1 -F_{1,\beta_2}\overline{W}_2\\
        -F_{2,\beta_1}\overline{W}_1+F_{1,\beta_1}\overline{W}_2
    \end{bmatrix} \quad \text{where} \quad \Lambda =   F_{1,\beta_1}F_{2,\beta_2} -  F_{1,\beta_2} F_{2,\beta_1} \label{eq:2paramsoln}
\end{equation}
In the rest of this section, we express $\sigbar$ in an orthonormal Cartesian basis \{$\nten{e}_x$,$\nten{e}_y$,$\nten{e}_z$\}. 

\subsection{Mohr-Coulomb strength surface}
\label{subsec:mc_results}

For the M-C surface, we note that $F_{i,\beta1} = \bar{\sigma}_{\texttt{max}i}$ and $F_{i,\beta2} = \bar{\sigma}_{\texttt{min}i}$ where $\bar{\sigma}_{\texttt{max}i}$ and $\bar{\sigma}_{\texttt{min}i}$ are the maximum and minimum principal values of $\sigbarsi$. We consider $\bar{\bm{\sigma}}_{\texttt{s}1}$ =  
  $(\sts/\sts)~ \nten{e}_x\otimes\nten{e}_x$ = $\nten{e}_x\otimes\nten{e}_x$,  so that ($\bar{\sigma}_{\texttt{max}1},\bar{\sigma}_{\texttt{min}1}$) = $(1,0)$. For $\bar{\bm{\sigma}}_{\texttt{s}2}$, we consider the following two different choices :\\

\noindent \textbf{Case (a) :} $\bar{\bm{\sigma}}_{\texttt{s}2}$ = $-\alpha_{\texttt{ct}}~ \nten{e}_x\otimes\nten{e}_x$  \quad \text{so that} \quad ($\bar{\sigma}_{\texttt{max}2},\bar{\sigma}_{\texttt{min}2}$) = $(0,-\alpha_{\texttt{ct}})$\vspace{0.2cm}

\noindent where  $\alpha_{\texttt{ct}} = {\scs}/{\sts} = 1/\alpha_{\texttt{tc}}$. \vspace{0.3cm} 

\noindent \textbf{Case (b) :} $\bar{\bm{\sigma}}_{\texttt{s}2}$ = $\alpha_{\texttt{st}} \nten{e}_x\otimes\nten{e}_y + \alpha_{\texttt{st}}\nten{e}_y\otimes\nten{e}_x$ \quad \text{so that} \quad ($\bar{\sigma}_{\texttt{max}2},\bar{\sigma}_{\texttt{min}2}$) = $(\alpha_{\texttt{st}},-\alpha_{\texttt{st}})$\vspace{0.2cm}

\noindent where  $\alpha_{\texttt{st}} = {\sss}/{\sts} = 1/\alpha_{\texttt{ts}}$.\\

\noindent Thus in Case (a), the phase field strength surface will exactly capture the unaxial tensile and compressive strengths for all $\eps$ whereas in Case (b), the uniaxial tensile and shear strengths will be captured. Note that using the above selection of $\sigbarsi$ along with \cref{eq:beta_strengthrelation} yields \cref{eq:MC_betas,eq:MC_betas2} for $\vv{\beta}$. We define the following quantities,
\begin{align}
 \overline{W}_{\texttt{ts}} = \overline{W}(\nten{e}_x\otimes\nten{e}_x), \ ~ \overline{W}_{\texttt{cs}} = \overline{W}(-\alpha_{\texttt{ct}}~ \nten{e}_x\otimes\nten{e}_x) ,\ ~\overline{W}_{\texttt{ss}} = \overline{W}(\alpha_{\texttt{st}} \nten{e}_x\otimes\nten{e}_y + \alpha_{\texttt{st}}\nten{e}_y\otimes\nten{e}_x) \label{eq:Wbars_sstscs}
\end{align}
so that using \cref{eq:MC_betas,eq:MC_betas2,eq:2paramsoln}, we arrive at the following solutions
\begin{align}
 &\text{Case (a) : }\quad   \beta_1^\eps = 1 - \frac{2 \Wbarts}{\bar{\omega}_\eps}, \quad \beta_2^\eps = -\alpha_{\texttt{tc}} \left(1 - \frac{2 \Wbarcs}{\bar{\omega}_\eps} \right)   \label{eq:Casea_sol_MC}\\
 &\text{Case (b) :}\quad \beta_1^\eps = 1 - \frac{2 \Wbarts}{\bar{\omega}_\eps}, \quad \beta_2^\eps = \left(1-\alpha_{\texttt{ts}}\right) - \frac{2}{\bar{\omega}_\eps}\left(\Wbarts -\alpha_{\texttt{ts}} \Wbarss \right) \label{eq:Caseb_sol_MC}
\end{align}
where we note that $\alpha_{\texttt{ts}} = 1 + \alpha_{\texttt{tc}}$.\\

\begin{figure}[h]
    \centering   \includegraphics[width=\textwidth]{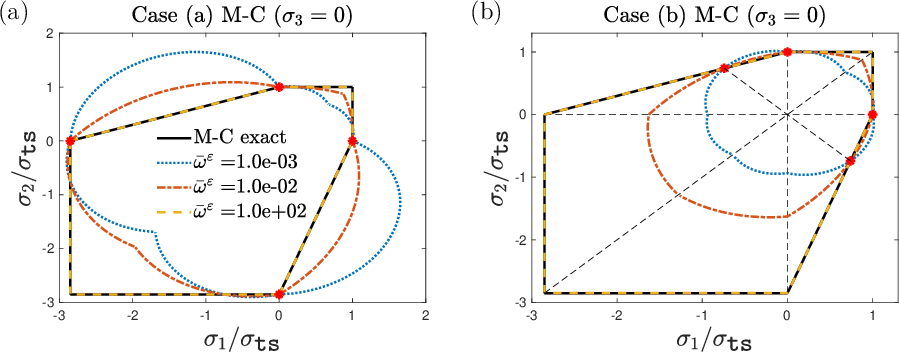}
    \caption{Comparison of phase field strength surface in  \cref{eq:final_general_ss_pf} using \cref{eq:Casea_sol_MC,eq:Caseb_sol_MC}, with the Mohr-Coulomb  material strength surface in \cref{eq:generalform_MC} for graphite parameters in \cref{eq:params_graphite}, shown in the principal plane space for plane stress loading. The points marked as red stars are the chosen values of $\sigbarsi$ that are captured exactly irrespective of $\eps$. (a) Case (a) outlined in \Cref{subsec:mc_results}, \cref{eq:Casea_sol_MC} wherein the uniaxial tensile and uniaxial compressive strengths are being captured exactly. (b) Case (b) outlined in \Cref{subsec:mc_results}, \cref{eq:Caseb_sol_MC} wherein the uniaxial tensile and shear strengths are being captured exactly. The legend is the same for both plots. Black dashed lines shown in (b) for visual guidance of the symmetries of the strength surface, the shear strength points lie on the line $y=-x$. }
    \label{fig:MC}
\end{figure}

To plot the phase field strength surfaces, we need to specify the form of the strain energy function. For our demonstration here, we consider a linear elastic strain energy function,
\begin{equation}
    \overline{W}(\sigbar) = \frac{1}{2} \left(\frac{\bar{J}_2}{\bar{\mu}} + \frac{\bar{I}_1^2}{9 \bar{K}}\right), \quad \bar{\mu} = \frac{\mu}{\sig^*}, \quad \bar{K} = \frac{K}{\sig^*} \label{eq:straineenergy}
\end{equation}
where $\mu$ is the shear modulus and $K$ is the bulk modulus. Using this in \cref{eq:Wbars_sstscs}, it is straightforward to show that
\begin{equation}
    2\overline{W}_{\texttt{ts}} = \frac{1}{3}\left(\frac{1}{\bar{\mu}} + \frac{1}{3 \bar{K}}\right), \quad \overline{W}_{\texttt{cs}} = \alpha_{\texttt{ct}}^2 \overline{W}_{\texttt{ts}}, \quad 2 \overline{W}_{\texttt{ss}} =  \left(\frac{\alpha_{\texttt{st}}^2}{\bar{\mu}}\right)
\end{equation}
We will now compare the exact material strength surface in \cref{eq:generalform_MC} with the phase field strength surface that is obtained by substituting \cref{eq:Casea_sol_MC,eq:Caseb_sol_MC} in \cref{eq:final_general_ss_pf}. The strength surface is a three dimensional surface in the space of principal stresses : $\sig_1$, $\sig_2$, $\sig_3$. We plot a cross-section of the principal stress space by considering plane stress loading such that $\sig_3 =0$. Results are shown in \Cref{fig:MC} by considering the following parameters for graphite used by \cite{KBFLP20} taken from \cite{goggin1967elastic,sato1987fracture}
\begin{equation}
    \mu = 4.3~\text{GPa},\quad K = 4.4~\text{GPa},\quad  \sts = 27~\text{MPa},\quad \scs= 77~\text{MPa} \label{eq:params_graphite}
\end{equation}\\

It can be seen that the phase field strength surfaces always coincide with the material strength surface irrespective of the value of $\bar{\omega}_\eps$ at the chosen strength locations marked in red star points\footnote{Note that there can be more than 2 marked points on the surface due to the symmetry of the strength space and the fact that we are plotting in principal stress space.} which verifies that our solution satisfies the finite $\eps$ match requirement(*). The chosen shear strength point in Case (b) will lie on the $y=-x$ line in the plane-stress principal stress space. Additionally, the phase field strength surfaces approach and coincide with the material strength surface for large $\bar{\omega}_\eps$ (small $\eps$) which means the solution is consistent. This verifies our solution for the M-C strength surface. Note that $\eps$ is simply a regularization length in the theory with no physical meaning ascribed to it such as attribution to process zone size as sometimes done in the classical phase field theories (to account for strength), consequently $\bar{\omega}_\eps$ is also simply a regularization parameter. We shall now look at the D-P case.

\subsection{Drucker-Prager strength surface}
\label{subsec:dp_results}

For the D-P surface, we note that $F_{i,\beta1} = \bar{I}_1(\sigbarsi)$ and $F_{i,\beta2} = \sqrt{\bar{J}_2(\sigbarsi)}$. Once again we consider $\bar{\bm{\sigma}}_{\texttt{s}1}$ =  $\nten{e}_x\otimes\nten{e}_x$, so that $F_{1,\beta_1} = 1$, $F_{1,\beta_2} = 1/\sqrt{3}$. We consider the following two choices $\bar{\bm{\sigma}}_{\texttt{s}2}$ :\\

\noindent 
\textbf{Case (a) :} $\bar{\bm{\sigma}}_{\texttt{s}2}$ = $\alpha_{\texttt{st}} \nten{e}_x\otimes\nten{e}_y + \alpha_{\texttt{st}}\nten{e}_y\otimes\nten{e}_x$ \quad \text{so that} \quad $F_{2,\beta_1} = 0$, $F_{2,\beta_2} = \alpha_{\texttt{st}}$  \vspace{0.2cm}

\noindent \textbf{Case (b) :} $\bar{\bm{\sigma}}_{\texttt{s}2}$ = $\alpha_{\texttt{bt}}~ \left(\nten{e}_x\otimes\nten{e}_x + \nten{e}_y\otimes\nten{e}_y \right)$ \quad \text{so that} \quad $F_{2,\beta_1} = 2 \alpha_{\texttt{bt}}$, $F_{2,\beta_2} = {\alpha_{\texttt{bt}}}/{\sqrt{3}}$ \vspace{0.3cm}

\noindent where  $\alpha_{\texttt{bt}} = {\sbs}/{\sts} = 1/\alpha_{\texttt{tb}}$.\\

\noindent Thus in Case (a), the phase field strength surface will exactly capture the unaxial tensile and shear strengths for all $\eps$ whereas in Case (b), the uniaxial tensile and tensile biaxial strengths will be captured. Using the above selection of $\sigbarsi$ along with \cref{eq:beta_strengthrelation} yields \cref{eq:DP_betas1,eq:DP_betas2} for $\vv{\beta}$. We note that solutions for choice of compressive strength or tensile hydrostatic strength for $\bar{\bm{\sigma}}_{\texttt{s}2}$ have been previously presented in the literature {\citep{KBFLP20,KRLP22,KKLP24}}, and hence our different constitutive choice of shear and tensile biaxial strengths instead, demonstrating in the process the ease of the driving force construction with our solution for any choice of $\sigbarsi$. Further, for many soft materials, the biaxial tensile strength may be more readily available than the hydrostatic tensile strength. \\

We define the following additional quantity,
\begin{align}
\overline{W}_{\texttt{bs}} = \overline{W}(\alpha_{\texttt{bt}} \nten{e}_x\otimes\nten{e}_y + \alpha_{\texttt{bt}}\nten{e}_y\otimes\nten{e}_x) \label{eq:Wbar_sbs}
\end{align}
which for our choice of strain energy in \cref{eq:straineenergy} simplifies as
\begin{equation}
   2\overline{W}_{\texttt{bs}} = \frac{\alpha_{\texttt{bt}}^2}{3}\left(\frac{1}{\bar{\mu}} + \frac{4}{3 \bar{K}}\right)
\end{equation}
Using \cref{eq:DP_betas1,eq:DP_betas2,eq:2paramsoln}, we arrive at the following solutions
\begin{align}
 &\text{Case (a) : }\quad   \beta_1^\eps = \left(1-\frac{\alpha_{\texttt{ts}}}{\sqrt{3}}\right) - \frac{2}{\bar{\omega}_\eps}\left( \Wbarts-\frac{\alpha_{\texttt{ts}}}{\sqrt{3}} \Wbarss \right), \quad \beta_2^\eps = \alpha_{\texttt{ts}} \left(1 - \frac{2 \Wbarss}{\bar{\omega}_\eps} \right)\label{eq:Casea_sol_DP}\\
 &\text{Case (b) :}\quad \beta_1^\eps = \left({\alpha_{\texttt{tb}}}-1\right) - \frac{2}{\bar{\omega}_\eps}\left( {\alpha_{\texttt{tb}}} \Wbarbs -\Wbarts\right), \quad \beta_2^\eps = \sqrt{3}\left(2-{\alpha_{\texttt{tb}}}\right) - \frac{2 \sqrt{3}}{\bar{\omega}_\eps}\left(2\Wbarts-\alpha_{\texttt{tb}} \Wbarbs\right) \label{eq:Caseb_sol_DP}
\end{align}
where we note that ${\alpha_{\texttt{ts}}} = \sqrt{3}\left(2 -{\alpha_{\texttt{tb}}}\right)$ and $2{\alpha_{\texttt{tb}}} = 3- {\alpha_{\texttt{tc}}}$.\\

\begin{figure}[h]
    \centering   \includegraphics[width=\textwidth]{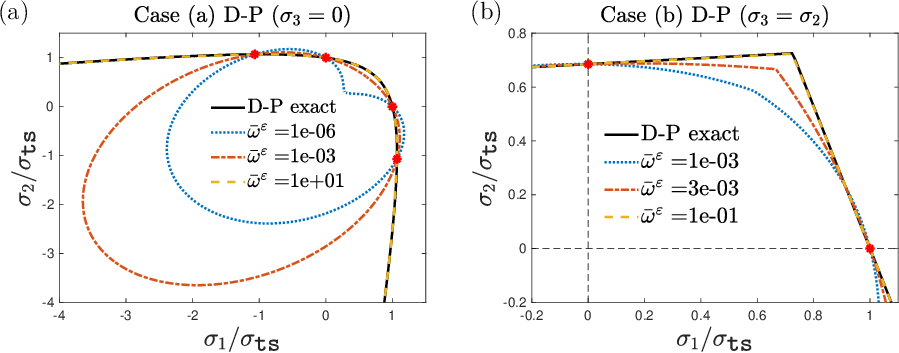}
    \caption{Comparison of phase field strength surface in  \cref{eq:final_general_ss_pf} using \cref{eq:Casea_sol_DP,eq:Caseb_sol_DP}, with the Drucker-Prager  material strength surface in \cref{eq:generalform_DP} for titania (TiO$_2$) parameters in \cref{eq:params_titania}, shown in the principal plane space for (a) plane stress loading for Case (a) outlined in \Cref{subsec:dp_results}, \cref{eq:Casea_sol_DP} and for (b) equibiaxial loading for Case (b) outlined in \Cref{subsec:dp_results}, \cref{eq:Caseb_sol_DP}. The points marked as red stars are the chosen values of $\sigbarsi$ that are captured exactly irrespective of $\eps$. (a) Uniaxial tensile and shear strengths are being captured exactly. (b) Uniaxial tensile and tensile biaxial strengths are being captured exactly.  Black dashed lines shown in (b) for visual guidance.}
    \label{fig:DP}
\end{figure}

We will now compare the exact material strength surface in \cref{eq:generalform_DP} with the phase field strength surface that is obtained by substituting \cref{eq:Casea_sol_DP,eq:Caseb_sol_DP} in \cref{eq:final_general_ss_pf}. Results are shown in \Cref{fig:DP} by considering the following parameters for titania (TiO$_2$) used by \cite{KBFLP20} taken from \cite{ely1972strength,iuga2007ab}
\begin{equation}
    \mu = 97~\text{GPa},\quad K = 198~\text{GPa},\quad  \sts = 100~\text{MPa},\quad \scs= 1232~\text{MPa} \label{eq:params_titania}
\end{equation} For Case (a), we plot in the plane stress principal stress space such that $\sig_3 =0$ and for Case (b), we plot in the equibiaxial stress principal stress space instead (for better visualization in this case) such that $\sig_3 = \sig_2$. Once again the consistency of the solution is seen and the chosen strength points are exactly captured irrespective of $\bar{\omega}_\eps$. This verifies our solution for the D-P strength surface as well.

Note that the D-P strength surface is not closed for $\alpha_{\texttt{tc}} > 3$ or  $\alpha_{\texttt{tc}} < 1/3$. For example, in the former case the material cannot fail under biaxial tension and in the latter case the material cannot fail under biaxial compression. For several materials, it is the case that $\alpha_{\texttt{tc}} < 1/3$ including the titania considered here. However note that the phase field strength surface is still closed for finite $\eps$. Thus the numerical phase field theory can potentially greatly underpredict failure in compression for finite $\eps$, which can deter predictive capability in modeling problems such as indentation. This is the motivation for potential correction terms such as in \cref{eq:I1_correction} \citep{KRLP22}. Note that since $\bar{I}_1$ is not negative for any of our $\sigbarsi$, the solution for coefficients $\vv{\beta^\eps}$ will remain unchanged by such a correction factor.

\section{Conclusions}

\label{sec:conclusions}

An explicit expression for a consistent crack nucleation driving force was derived for the modified phase field theories given a general strength surface linear in material parameter coefficients.  In the limit of vanishing regularization length $\eps$, it leads to the prediction of the material strength surface exactly and for finite $\eps$ it predicts certain chosen strength locations on the material strengh surface exactly for all $\eps$. Results for the Mohr-Coulomb strength surface are a first and can facilitate application of the theory to brittle materials such as concrete and rocks as well as potential extension the theory to predict crack nucleation in ductile fracture \citep{BaiWierzbicki2010}. Efforts in these directions are underway. The results for a general strength surface can facilitate potential future integration of other phenomenological \citep{peng2021new} as well as micromechanics based ductile failure criteria \citep{keralavarma2020ductile,keralavarma2016criterion} (that are grounded in underlying micromechanical phenomena and applicable in more general non-proportional loading scenarios) into the phase field modeling of fracture, similar to previous integration into the extended finite element method (XFEM) using cohesive segments \citep{sidharth2022crack,nikhil2024application}. Results for the Drucker-Prager surface are new for the constitutive choice of strength locations chosen to be captured exactly for all $\eps$. For crack nucleation, the emergence of an effective toughness term dependent on the stress is shown that is zero at strength based initiation of failure and attains a maximum value in the cracked state. Potential parallels to cohesive zone models can be explored which might illuminate other ways to possibly calibrate $\de$ to marry the strength with fracture toughness $G_c$. 

\section*{Acknowledgements}

\noindent The author would like to acknowledge insightful discussions with Aditya Kumar at the Georgia Institute of Technology.

%\bibliographystyle{elsarticle-num-names}
%\bibliography{ref}

\end{document}